\documentclass[aps,pra,amsmath,showpacs,twocolumn,floatfix]{revtex4}
\pdfoutput=1

\usepackage{hyperref}
\usepackage{graphicx}
\usepackage{natbib}
\usepackage{mhchem}

\graphicspath{{figs/}}

\renewcommand{\eqref}[1]{Eq.~(\ref{#1})}

\begin{document}

\title{Production of translationally cold barium monohalide ions}
\date{\today}
\author{M.~V.~DePalatis}
\author{M.~S.~Chapman}
\affiliation{School of Physics, Georgia Institute of Technology,
  Atlanta, Georgia 30332-0430}

\begin{abstract}
  We have produced sympathetically cooled barium monohalide ions
  \ce{BaX+} (\ce{X} = \ce{F}, \ce{Cl}, \ce{Br}) by reacting trapped,
  laser cooled \ce{Ba+} ions with room temperature gas phase neutral
  halogen-containing molecules. Reaction rates for two of these
  (\ce{SF6} and \ce{CH3Cl}) have been measured and are in agreement
  with classical models. \ce{BaX+} ions are promising candidates for
  cooling to the rovibrational ground state, and our method presents a
  straightforward way to produce these polar molecular ions.
\end{abstract}


\maketitle

In recent years, considerable progress has been made in extending
laser cooling techniques to both neutral molecules
\cite{Shuman2010,Manai2012,Stuhl2012,Hummon2013} and molecular ions
\cite{Staanum2010,Nguyen2011,Bressel2012}. Cold molecular ions are of
particular interest for the study of chemical reactions in the quantum
regime \cite{Willitsch2008a}, electron electric dipole moment
searches, \cite{Meyer2008,Cossel2012}, testing for time variation of
fundamental constants \cite{Flambaum2007}, and for use in cavity QED
experiments \cite{Schuster2011}. One particularly promising class of
molecular ions for cooling and precision spectroscopy are alkaline
earth monohalide ions which consist of two closed-shell atomic
ions. These molecules, among others, can be cooled to the
rovibrational ground state via collisions with ultracold neutral atoms
\cite{Hudson2009,Rellergert2013}. One method for producing such ions,
employed in Refs.~\cite{Chen2011,Schowalter2012,Rellergert2013}, is
ablating an appropriate target. Several studies have produced other
alkaline earth monohalides such as \ce{CaF+}
\cite{Harvey1997,Willitsch2008,Gingell2010}. Additionally, \ce{Ba+} is
known to react with \ce{HCl} to form \ce{BaCl+} \cite{Schowalter2012},
but in this case, the presence of \ce{HCl} was an unintended byproduct
formed by reaction of background gases and ablation products from a
\ce{BaCl2} target.

\begin{figure}
  \centering
  \includegraphics{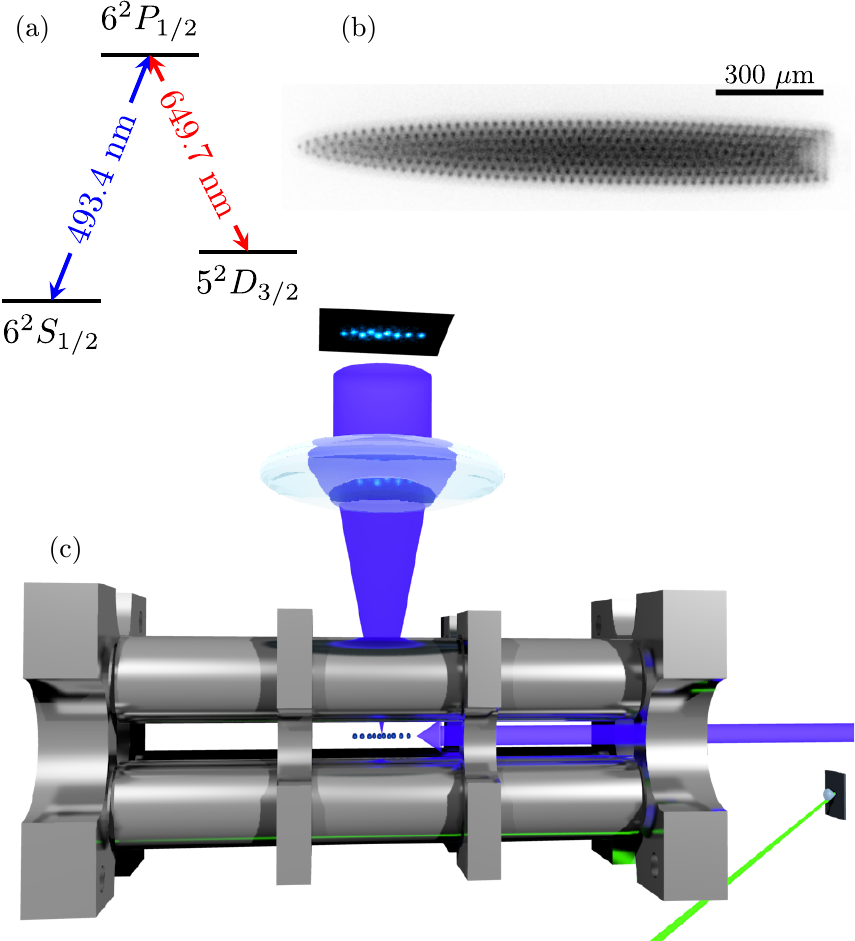}
  \caption{(Color online) (a) \ce{Ba+} energy levels and optical
    transitions. (b) A typical Coulomb crystal containing several
    hundred \ce{Ba+} ions. Radiation pressure forces \ce{^{138}Ba+}
    ions in the direction of laser propagation (to the left) and the
    remaining isotopes collect on the other end (right). (c) Schematic
    drawing of the experimental setup. \ce{Ba+} ions are loaded via
    laser ablation (diagonal beam) and cooled axially (horizontal
    beam). Fluorescence images are obtained with a NA = 0.34
    achromatic lens and magnified by a factor of 4.}
  \label{fig:ExpSetup}
\end{figure}

Here, we produce translationally cold \ce{BaX+} ions (\ce{X} = \ce{F},
\ce{Cl}, \ce{Br}) through reactions with neutral molecules at room
temperature and trapped \ce{Ba+} ions at mK temperatures. We determine
the reaction rate constants for the production of \ce{BaF+} and
\ce{BaCl+}, and we utilize both non-destructive motional resonance
coupling and mass-selective ejection to verify the reaction
products. Producing \ce{BaX+} ions in this way is simple and allows
for the study of different barium monohalide species without
significant changes in the experimental setup.

The experiment begins with Coulomb crystals of \ce{Ba+} ions formed in
a linear Paul trap with radius $r_0 = 3.18$ mm and RF frequency
$\Omega = 2\pi \times 2.7$ MHz with RF applied to all four electrodes
(Fig.~\ref{fig:ExpSetup}). Typical peak-to-peak RF voltages are $V
\approx$ 100--400 V for trapping, and up to $\sim 2$ kV during
loading. The barium ions are loaded by ablating a barium metal target
in a similar manner as described previously \cite{Churchill2011}. Ions
are confined axially with end caps separated by 25 mm. Typical end cap
voltages are $U_{\text{EC}} \approx$ 100--300 V. Cooling ($\lambda =
493$ nm) and repumper ($\lambda = 650$ nm) beams are introduced
axially from one end of the trap. This causes an axial sorting of
barium isotopes visible as an apparent asymmetry in the Coulomb
crystal structure as shown in Fig.~\ref{fig:ExpSetup}. We typically
tune the cooling and repumper beams such that they are red detuned for
maximum fluorescence of $^{138}$Ba$^+$ since it is the most abundant
isotope and therefore maximizes the fluorescence signal. The lasers
are separately stabilized to temperature controlled cavities that are
in turn locked to a 780 nm laser referenced to a \ce{Rb} vapor cell
(see Ref.~\cite{Rohde2010} for a similar locking scheme).

\begin{figure}
  \centering
  \includegraphics{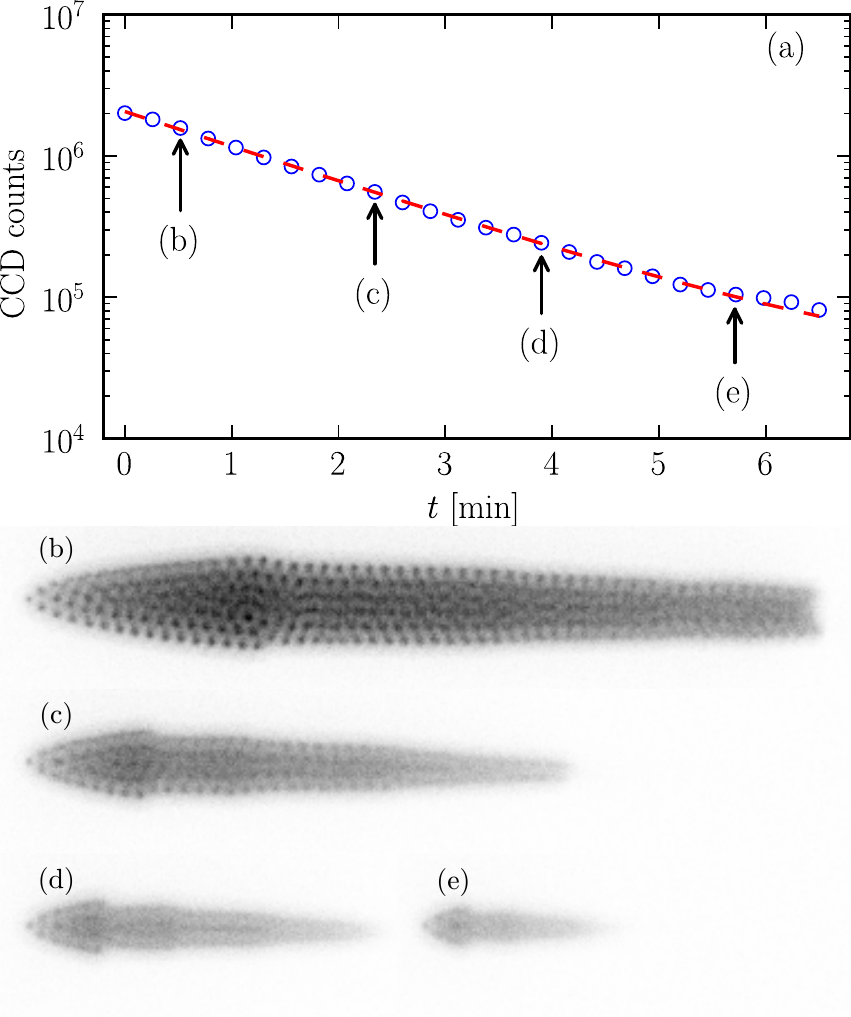}
  \caption{(Color online) (a) A typical loss rate measurement between
    \ce{Ba+} and \ce{SF6} at a partial pressure of $2.8 \times
    10^{-10}$ torr using total fluorescence. The dashed line is a
    nonlinear least squares fit to an exponential decay from which the
    loss rate is extracted. (b)--(e) Snapshots of the Coulomb crystal
    at each point indicated in (a).}
  \label{fig:SF6LossRate}
\end{figure}

In order to create \ce{BaX+} molecular ions, the reactants are leaked
into the vacuum chamber at partial pressures of up to $10^{-9}$
torr. At these pressures, the laser cooled ions remain in an ordered
state. The reactions result in the loss of \ce{Ba+} ions and so a loss
rate can be determined by measuring the total fluorescence over time
(Fig.~\ref{fig:SF6LossRate}). The loss rates can then be transformed
into reaction rate constants by measuring the loss rate at different
pressures and extracting the slope of a linear fit to these data. We
utilize \ce{SF6}, \ce{CH3Cl}, and \ce{Br2} as reactants. \ce{SF6} and
\ce{CH3Cl} are in the gas phase at room temperature and are stored in
a reservoir of about 500 mL at pressures of slightly more than 1 atm
which is connected to the inlet of the leak valve. Bromine is a liquid
at room temperature, but has a sufficiently high vapor pressure of
\ce{Br2} ($\sim 185$ torr \cite{Lide2007}) to use as a gas phase
reactant. A few mL of liquid bromine is added to a small reservoir at
rough vacuum which is connected to the leak valve inlet.

The simplest model for characterizing the rate of reactions between
ions and neutral, nonpolar molecules is the Langevin model
\cite{Gioumousis1958}. Assuming spherical symmetry and given the
charge $Ze$ on the ion and polarizability $\alpha$ of the neutral
molecule, the Langevin reaction rate constant can be expressed in
Gaussian-cgs units as

\begin{equation}
  \label{eq:LangevinRate}
  k_L = 2\pi Ze \sqrt{\frac{\alpha}{\mu}},
\end{equation}
where $\mu$ is the reduced mass. Typical rates calculated from this
model are of order $10^{-9}$ cm$^3$ s$^{-1}$. The Langevin model is
not comprehensive, but nevertheless serves as a useful estimate of
reaction rates and many nonpolar molecules do indeed react at or near
the predicted rate. For polar molecules, the Langevin rate must be
corrected by taking into account the dipole moment $d$. The corrected
rate constant is expressed as \cite{Troe1985}

\begin{equation}
  \label{eq:ADORate}
  k_c = k_L + ck_D = 2\pi Ze \left(
    \sqrt{\frac{\alpha}{\mu}} + cd \sqrt{\frac{2}{\pi \mu k_B T}} \right),
\end{equation}
where $0 < c < 1$ and $k_D$ is the component of the rate constant
associated with a ``locked in'' dipole. There are several methods for
estimating $c$ including the average dipole orientation theory
\cite{Su1973}, but to lowest order, by setting $c = 1$,
\eqref{eq:ADORate} can be used to estimate an upper bound on the
reaction rate.

In order to translate loss rate measurements into a reaction rate
constant, we measure the loss rate of \ce{Ba+} ions at several
different reactant gas pressures. The measured reaction rate constant
$k$ is then the slope of a linear fit to the loss rate versus pressure
data. We use a Varian UHV-24 ion gauge to measure pressures and apply
appropriate correction factors $\epsilon$ in order to obtain a reading
more suitable for the reactant introduced. This gives us corrected
rate constants $k_\epsilon = \epsilon k$. Ultimately, since the ion
gauge was not calibrated, the dominant source of error lies with the
pressure measurements and thus the measured values $k_\epsilon$ are
accurate only within a factor of 2 or 3.

\begin{figure}
  \centering
  \includegraphics{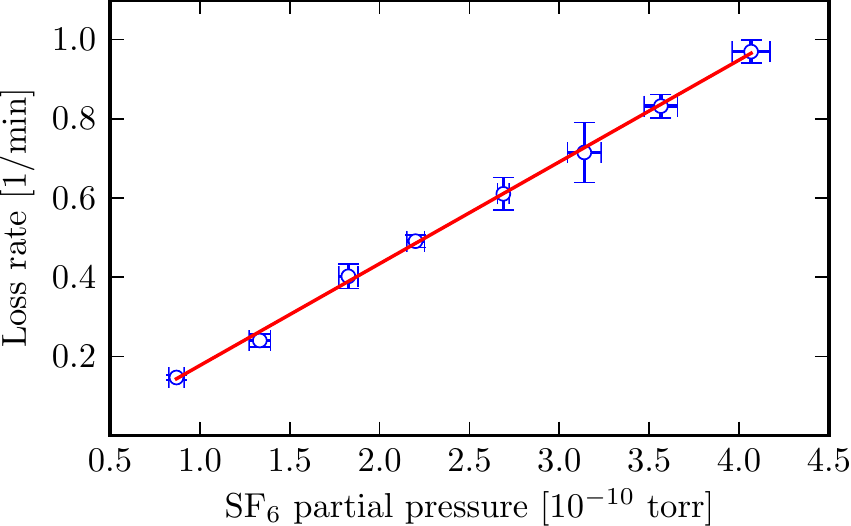}
  \caption{(Color online) \ce{Ba+} + \ce{SF6} reaction rate
    measurements. Each point represents the average measured loss rate
    at a given target partial pressure. Vertical error bars represent
    the standard deviation of the measured loss rates and horizontal
    error bars indicate the standard deviation of the pressure
    recorded by the ion gauge for each measurement. The solid line is
    a linear fit to the data. Partial pressures are corrected using
    the ion gauge correction factor $\epsilon = 2.3$.}
  \label{fig:SF6RxnRate}
\end{figure}

To form \ce{BaF+}, we utilize \ce{SF6}. The expected reaction is

\begin{equation}
  \label{eq:BaSF6}
  \ce{Ba^+ + SF6 -> BaF^+ + SF_5},
\end{equation}
which from known thermochemistry is exothermic by 2.8 eV.  In order to
determine the reaction rate constant $k$ between \ce{Ba+} and
\ce{SF6}, a series of measurements of the \ce{Ba+} loss rate were made
at several different partial pressures of \ce{SF6}. A linear fit of
the measured loss rates versus partial pressure then yields the
reaction constant (Fig.~\ref{fig:SF6RxnRate}) which is listed in Table
\ref{tab:ReactionRates}. We find that the reaction between \ce{Ba+}
and \ce{SF6} proceeds in good agreement with the Langevin model.

\begin{table}
  \centering
  \caption{Reaction rate constants between \ce{Ba+} and neutral 
    reactants used in this work in units of $10^{-9}$ cm$^3$
    s$^{-1}$. Reaction rate constants $k$ using uncorrected ion
    gauge pressure readings are multiplied by ion gauge correction
    factors $\epsilon$ to obtain the corrected reaction rate constants
    $k_\epsilon = \epsilon k$. Ion gauge correction factors are
    obtained from \cite{Varian2004}. Also listed are the theoretical
    Langevin rate constant ($k_L$), dipole correction term ($k_D$),
    and upper bound corrected Langevin rate constant ($k_c \leq k_L +
    k_D$) using polarizabilities from
    \cite{Karpas1989,Olney1997,Maroulis1997,Maroulis1997a}
    and dipole moment from \cite{Johnson2011}.}
  \begin{ruledtabular}
    \begin{tabular}{ccccccc}
      Reactant & $k$ & $\epsilon$ & $k_\epsilon$ & $k_L$ & $k_D$ & $k_c$ \\ \hline
      \ce{SF6} & 0.58 & 2.3 & 1.3 & 0.59 & --- & --- \\
      \ce{CH3Cl} & 0.63 & 2.6 & 1.6 & 0.82 & 2.87 & $\leq 3.69$ \\
      \ce{Br2} & --- & 3.8 & --- & 0.70 & --- & --- \\
      \ce{I2} & --- & 5.4 & --- & 0.80 & --- & --- \\
    \end{tabular}
    \label{tab:ReactionRates}
  \end{ruledtabular}
\end{table}

\begin{figure}
  \centering
  \includegraphics{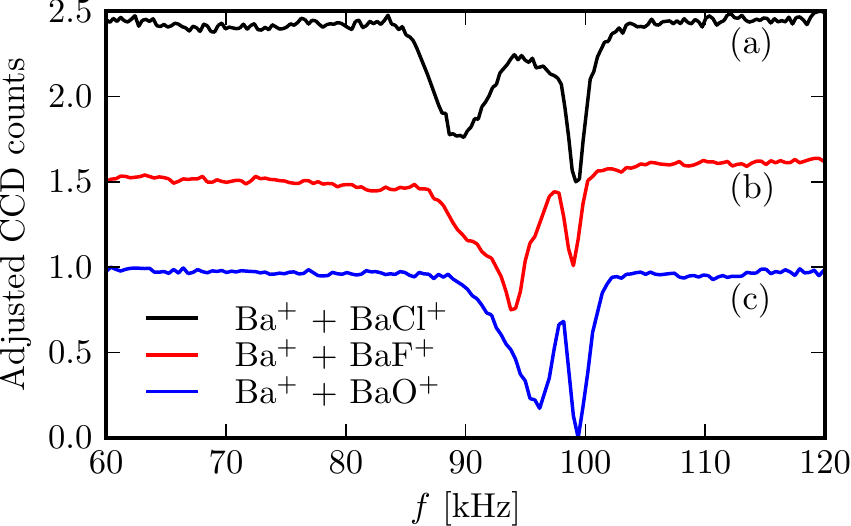}
  \caption{(Color online) AC frequency sweeps after \ce{Ba+} Coulomb
    crystals react with (a) \ce{CH3Cl}, (b) \ce{SF6} and (c)
    \ce{O2}. CCD counts are scaled and offset for clarity. In each
    case, the broader resonance at lower frequencies is due to the
    heavier sympathetically cooled product ions while the narrower
    resonance at higher frequencies is due to the laser cooled
    \ce{Ba+} ions.}
  \label{fig:SecularCompare}
\end{figure}

Energy considerations imply that \eqref{eq:BaSF6} should be the
dominant reaction between \ce{Ba+} and \ce{SF6}, but to verify this,
we performed mass spectroscopic measurements on reaction products. One
common method is to excite motional resonances of the trapped ions by
applying an additional AC voltage to the trap electrodes. Ions are
heated when in resonance with this applied voltage which results an
increase of temperature for all trapped ions through their mutual
Coulomb repulsion. This increased temperature changes the laser cooled
ion fluorescence and thus can be used to determine the ion masses
\cite{Roth2005,Roth2007,Landa2012}. For a single ion, axial and radial
frequencies are given respectively by $\omega_z = (2Ze\kappa
U_{\text{EC}}/mz_0^2)^{1/2}$ and $\omega_r = (\omega_0^2 -
\omega_z^2/2)^{1/2}$, where $\omega_0 = ZeV/\sqrt{2}mr_0^2\Omega$,
$\kappa$ is a unitless geometric constant, and $z_0$ is half the
distance between the end caps. However, co-trapping ions of different
species is well known to introduce shifts away from these frequencies
which complicates attempts to use the AC resonance frequencies to
determine the masses of the trap contents with good precision
\cite{Baba2002}. Furthermore, the preceding expressions are derived
assuming the trapping pseudopotential is quadratic in each
dimension. This is generally a good approximation in the radial
direction, but due to the geometry of our trap, the axial potential
deviates from a quadratic approximation significantly for large
Coulomb crystals. Nevertheless, frequency sweeps are still useful as
they are nondestructive and provide information about the mass of the
sympathetically cooled ions relative to that of the laser cooled
ions. We drive radial modes by applying AC signals of several hundred
mV to one of the trap electrodes. Typical results from frequency
sweeps of Coulomb crystals composed of \ce{Ba+} and reaction products
(\ce{A+}) are presented in Fig.~\ref{fig:SecularCompare}.

It should be noted that the mass of \ce{BaF+} ($\approx 157$ amu) is
very near the mass of \ce{SF6+} ($\approx 146$ amu). Given the known
frequency shift effects and fairly low mass resolution, frequency
sweeps alone are not sufficient to preclude the possibility that the
reaction between \ce{Ba+} and \ce{SF6} is simple charge
exchange. However, the ionization energy of \ce{SF6} is known to be
$>15$ eV compared to 5.2 eV for \ce{Ba} \cite{Lias2011,Karlsson1999},
making such a reaction energetically unfavorable. Furthermore,
\ce{SF6+} is expected to rapidly decay into \ce{SF5+} which is lighter
than \ce{^{138}Ba+} by more than 10 amu \cite{Tachikawa2000}. Lighter
sympathetically cooled ions are more tightly confined at a given RF
voltage, and therefore form a light core at the center of the Coulomb
crystal. Likewise, heavier sympathetically cooled ions form shells on
the exterior of the Coulomb crystal. Since we only observe the latter
crystal structures, we conclude all reaction products are heavier than
\ce{Ba+}.

\begin{figure}
  \centering
  \includegraphics{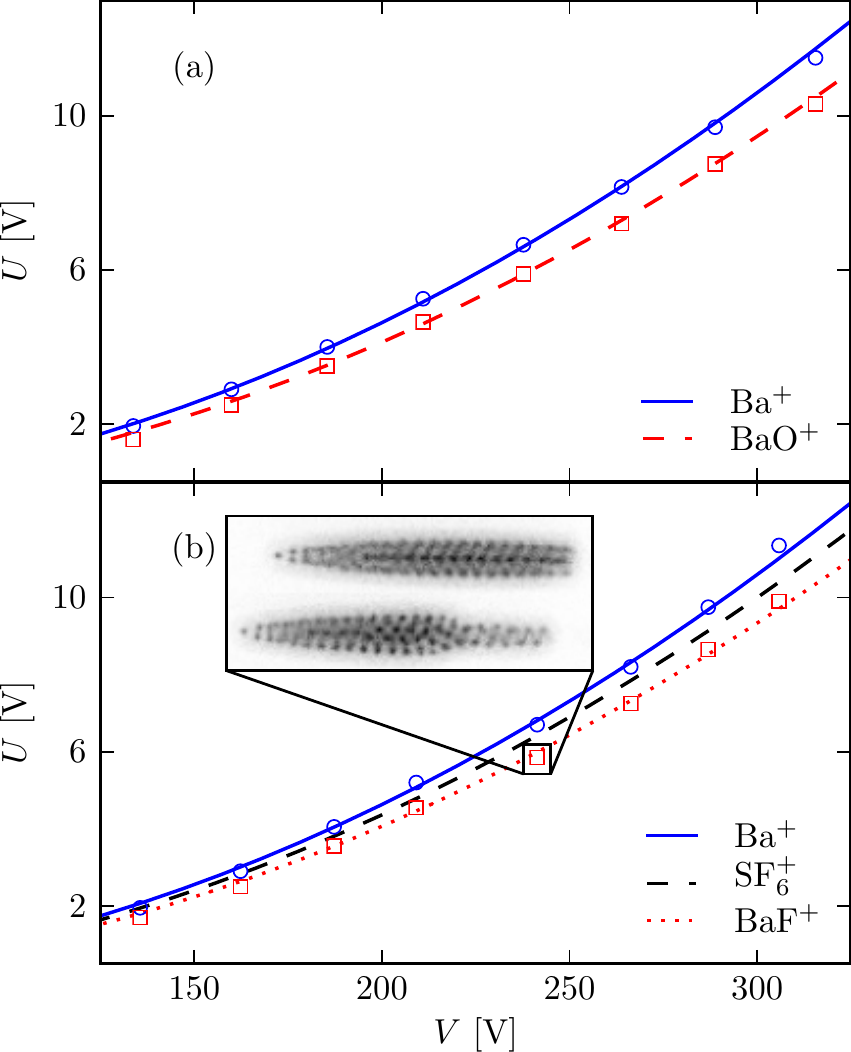}
  \caption{(Color online) Analysis of products of the reaction between
    (a) \ce{Ba+} and \ce{O2} and (b) \ce{Ba+} and \ce{SF6}. By
    carefully applying DC offsets to each RF electrode, heavy reaction
    products can be ejected from the trap without losing any
    \ce{Ba+}. The inset images of (b) show crystals before (bottom)
    and after (top) applying voltages to eject \ce{BaF+}. Note both a
    change in crystal structure and an overall shift towards the trap
    center (to the right). Solid curves represent the theoretical
    stability region boundaries.}
  \label{fig:Destructive}
\end{figure}

To further rule out the possibility of charge exchange between
\ce{Ba+} and \ce{SF6}, we also performed a series of destructive
measurements utilizing the Mathieu stability parameters $q =
2ZeV/mr_0^2\Omega^2$ and $a = 4ZeU/mr_0^2\Omega^2$.
By applying DC offset voltages $\pm U/2$ to each RF electrode, a
portion of the $q$-$a$ stability boundary for heavy, sympathetically
cooled \ce{A+} ions can be determined by watching for a change in the
crystal structure and a contraction of the crystal which appears as an
overall shift of the fluorescing ions toward the center of the
trap. These changes in the crystal indicate the ejection of the
heavier ions thereby allowing for determination of their
mass. \ce{Ba+} ions are first loaded into the trap, then allowed to
react with \ce{SF6} for several seconds until the crystal structure
clearly indicates the presence of heavier sympathetically cooled
ions. The RF voltage is then brought to a particular value of $V$ and
DC offsets are pulsed with increasing voltage until the crystal
structure changes and shifts towards the center of the trap,
indicating the ejection of \ce{A+} ions. This value of $U$ is recorded
and the DC voltage is increased further in order to find the value of
$U$ at which \ce{Ba+} is ejected. Overlaying these $U$, $V$ pairs on
top of the theoretical stability boundaries then gives us a more
precise determination of the product mass than do AC frequency
sweeps. Because the probe for measuring RF voltage is not well
calibrated, the probe reading is multiplied by a scaling factor which
is chosen by fitting the $U$, $V$ pairs for \ce{Ba+} to its
theoretical stability boundary. In order to test this method, we
performed similar measurements on Coulomb crystals consisting of
\ce{Ba+} and \ce{BaO+} formed by leaking \ce{O2} into the
chamber. This utilizes the endothermic reaction $\ce{Ba+ + O2 -> BaO+
  + O}$ which has been studied extensively elsewhere
\cite{Murad1982,Roth2008}. The results of these measurements following
both reactions \ce{Ba^+ + O2} and \ce{Ba^+ + SF6} are shown in
Fig.~\ref{fig:Destructive}. Given these results and the previously
discussed energy considerations, we conclude that we are in fact
producing \ce{BaF+} by the reaction \eqref{eq:BaSF6}.

\begin{figure}
  \centering
  \includegraphics{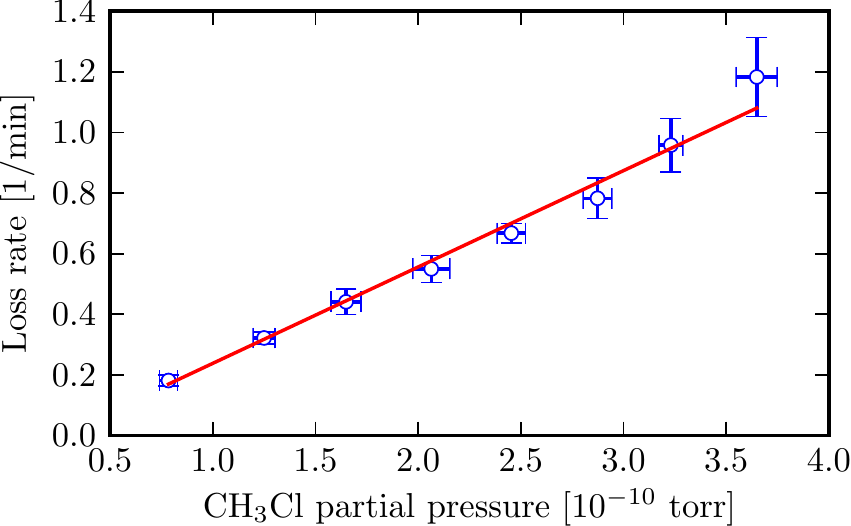}
  \caption{(Color online) \ce{Ba+ + CH3Cl} reaction rate
    measurements. Partial pressures are corrected using the ion gauge
    correction factor $\epsilon = 2.6$.}
  \label{fig:CH3ClRxnRate}
\end{figure}

For the production of \ce{BaCl+}, \ce{Ba+} was reacted with \ce{CH3Cl}
via the reaction

\begin{equation}
  \label{eq:BaCH3Cl}
  \ce{Ba+ + CH3Cl -> BaCl+ + CH3}.
\end{equation}
The rate constant was measured in the same manner as described above
for reactions with \ce{SF6}. The results of the reaction rate
measurement are shown in Fig.~\ref{fig:CH3ClRxnRate} and the
associated rate constant compared with theory is listed in
Table~\ref{tab:ReactionRates}. As with the \ce{Ba+ + SF6} reaction,
our measurements are in good agreement with the dipole-corrected
Langevin model.

The AC frequency sweep in Fig.~\ref{fig:SecularCompare} shows that
\eqref{eq:BaCH3Cl} is likely the dominant reaction between \ce{Ba+}
and \ce{CH3Cl}. However, this data alone is not sufficient to entirely
rule out some production of \ce{BaH+} via the reaction

\begin{equation}
  \label{eq:BaHRxn}
  \ce{Ba+ + CH3Cl -> BaH+ + CH2Cl}
\end{equation}
since the mass spectroscopic techniques here are not sufficient to
resolve mass differences of 1 amu \footnote{We were unable to locate
  sufficient data in the literature to estimate the reaction enthalpy
  for the reaction $\text{Ba}^+ + \text{CH}_3\text{Cl} \rightarrow
  \text{BaCH}_3^+ + \text{Cl}$ and therefore determine whether or not
  it would be energetically allowable. However, it is sufficiently
  heavier than \ce{Ba+} to be resolvable with AC frequency
  sweeps. Since no such resonances were observed, this reaction can be
  ruled out.}. The reaction (\ref{eq:BaHRxn}) is estimated to be
endothermic by 2.3 eV which would preclude it from occurring with
\ce{Ba+} in the ground state \cite{Lide2007}. However, \ce{Ba+} ions
in the $6^2P_{1/2}$ excited state are able to overcome this energy
barrier such that \eqref{eq:BaHRxn} becomes exothermic by 0.2 eV. This
implies that there is likely some amount of \ce{BaH+} being
produced. Future studies with improved mass resolution via
time-of-flight mass spectrometry \cite{Schowalter2012} or high
precision non-destructive techniques \cite{Drewsen2004} will be
necessary to fully analyze reaction products.

\begin{figure}
  \centering
  \includegraphics{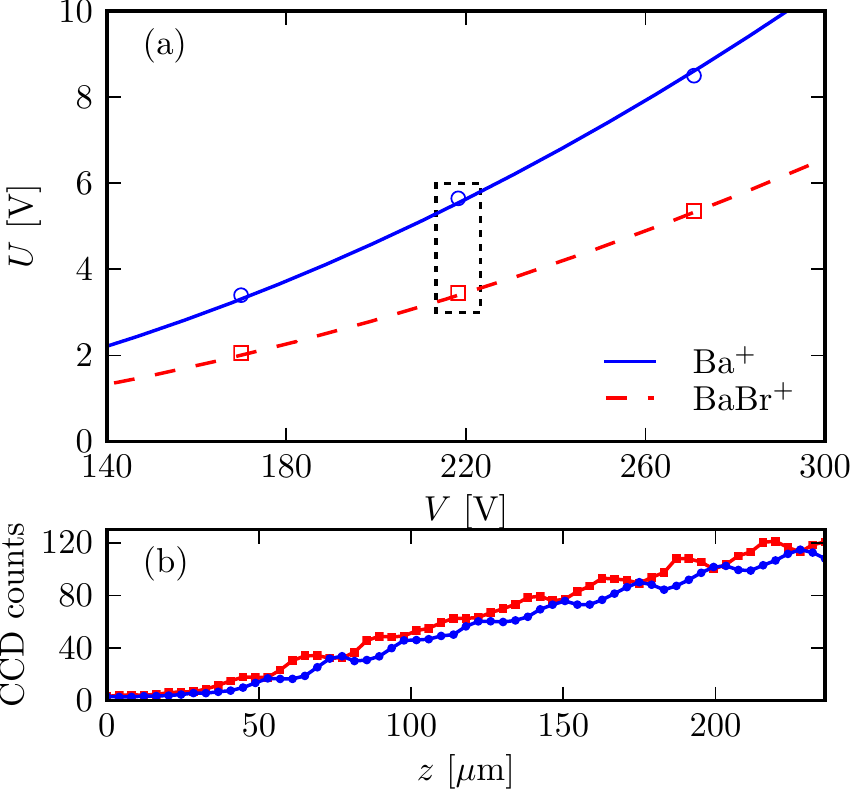}
  \caption{(Color online) (a) Destructive mass determination following
    reactions between \ce{Ba+} and \ce{Br2}. (b) Integrated CCD counts
    along the center and near one end of the Coulomb crystals
    indicated with the dotted box in (a) both before (curve with
    squares) and after (curve with circles) applying a $U$ pulse to
    eject \ce{BaBr+} ions. The shift to the right indicates the
    successful ejection of heavy ions.}
  \label{fig:BaBrDestructive}
\end{figure}

\ce{BaBr+} is produced through the reaction

\begin{equation}
  \label{eq:BaBr2Rxn}
  \ce{Ba+ + Br2 -> BaBr+ + Br},
\end{equation}
which is exothermic by about 2.2 eV. In order to get enough \ce{Br2}
in the vicinity of the trapped \ce{Ba+} ions, it was necessary to
increase the leak rate considerably compared to what was required for
\ce{SF6} and \ce{CH3Cl}. This allowed other contaminants such as
\ce{O2}, \ce{CO2}, and \ce{H2O} into the chamber simultaneously which
also react with \ce{Ba+}. However, the first two reactions are
endothermic with the ground state \cite{Roth2008}, and the last is
barely exothermic \cite{Murad1982}. Thus by blocking the cooling
beams, primarily \ce{BaBr+} is produced. Measuring loss rates was not
possible because of this, but it appeared to be much slower than with
the other reactants. We suspect this was in part due to the liquid
nature of \ce{Br2} at room temperature combined with the fact that
there was no direct line of sight between the leak valve and the
trapped ions. Due to the slow apparent loss rate, producing enough
\ce{BaBr+} to obtain a good AC frequency sweep mass spectrum was not
practical, so destructive measurements as in
Fig.~\ref{fig:Destructive} were performed and are shown in
Fig.~\ref{fig:BaBrDestructive}. Because the number of product ions was
very small compared to the number of \ce{Ba+} ions, ejection of
reactants is noted by watching for a shift in the outermost ion
towards the center rather than watching for a large change in crystal
structure.

We also attempted to produce \ce{BaI+} in a similar manner using the
reaction

\begin{equation}
  \label{eq:BaI2Rxn}
  \ce{Ba+ + I2 -> BaI + I}
\end{equation}
which is also exothermic by about 2.2 eV. \ce{I2} is a solid at room
temperature with a vapor pressure of around 300 mtorr. A sample of a
few grams of \ce{I2} was placed in a small reservoir evacuated to
rough vacuum before being connected to the leak valve inlet. No
\ce{BaI+} production was observed, presumably for similar reasons as
mentioned above. Future studies with a more optimal chamber geometry
could allow for study of reaction (\ref{eq:BaI2Rxn}) at room
temperature.

In summary, we have demonstrated a simple method for the production of
translationally cold \ce{BaF+}, \ce{BaCl+}, and \ce{BaBr+} by reacting
\ce{Ba+} ions with \ce{SF6}, \ce{CH3Cl}, and \ce{Br2}. Reaction rate
constants between \ce{Ba+} and \ce{SF6} and \ce{CH3Cl} were measured
and found to be in good agreement with classical predictions within
the limitations of pressure measurements. With some changes to the
vacuum chamber, similar rate constant measurements could be made for
the reactions of \ce{Ba+} with \ce{I2} and \ce{Br2}.

\acknowledgments We thank Michael Schatz's lab for the use of
\ce{SF6}. We also gratefully acknowledge useful discussions with
Richard Darst, James Goeders, and Ncamiso Khanyile. This work was
supported by the National Science Foundation and the Office of Naval
Research.

\bibliographystyle{apsrev}
\bibliography{BaRxn}

\begin{thebibliography}{41}
\expandafter\ifx\csname natexlab\endcsname\relax\def\natexlab#1{#1}\fi
\expandafter\ifx\csname bibnamefont\endcsname\relax
  \def\bibnamefont#1{#1}\fi
\expandafter\ifx\csname bibfnamefont\endcsname\relax
  \def\bibfnamefont#1{#1}\fi
\expandafter\ifx\csname citenamefont\endcsname\relax
  \def\citenamefont#1{#1}\fi
\expandafter\ifx\csname url\endcsname\relax
  \def\url#1{\texttt{#1}}\fi
\expandafter\ifx\csname urlprefix\endcsname\relax\def\urlprefix{URL }\fi
\providecommand{\bibinfo}[2]{#2}
\providecommand{\eprint}[2][]{\url{#2}}

\bibitem[{\citenamefont{Shuman et~al.}(2010)\citenamefont{Shuman, Barry, and
  {DeMille}}}]{Shuman2010}
\bibinfo{author}{\bibfnamefont{E.~S.} \bibnamefont{Shuman}},
  \bibinfo{author}{\bibfnamefont{J.~F.} \bibnamefont{Barry}}, \bibnamefont{and}
  \bibinfo{author}{\bibfnamefont{D.}~\bibnamefont{{DeMille}}},
  \bibinfo{journal}{Nature} \textbf{\bibinfo{volume}{467}},
  \bibinfo{pages}{820} (\bibinfo{year}{2010}).

\bibitem[{\citenamefont{Manai et~al.}(2012)\citenamefont{Manai, Horchani,
  Lignier, Pillet, Comparat, Fioretti, and Allegrini}}]{Manai2012}
\bibinfo{author}{\bibfnamefont{I.}~\bibnamefont{Manai}},
  \bibinfo{author}{\bibfnamefont{R.}~\bibnamefont{Horchani}},
  \bibinfo{author}{\bibfnamefont{H.}~\bibnamefont{Lignier}},
  \bibinfo{author}{\bibfnamefont{P.}~\bibnamefont{Pillet}},
  \bibinfo{author}{\bibfnamefont{D.}~\bibnamefont{Comparat}},
  \bibinfo{author}{\bibfnamefont{A.}~\bibnamefont{Fioretti}}, \bibnamefont{and}
  \bibinfo{author}{\bibfnamefont{M.}~\bibnamefont{Allegrini}},
  \bibinfo{journal}{Phys. Rev. Lett.} \textbf{\bibinfo{volume}{109}},
  \bibinfo{pages}{183001} (\bibinfo{year}{2012}).

\bibitem[{\citenamefont{Stuhl et~al.}(2012)\citenamefont{Stuhl, Hummon, Yeo,
  Quéméner, Bohn, and Ye}}]{Stuhl2012}
\bibinfo{author}{\bibfnamefont{B.~K.} \bibnamefont{Stuhl}},
  \bibinfo{author}{\bibfnamefont{M.~T.} \bibnamefont{Hummon}},
  \bibinfo{author}{\bibfnamefont{M.}~\bibnamefont{Yeo}},
  \bibinfo{author}{\bibfnamefont{G.}~\bibnamefont{Quéméner}},
  \bibinfo{author}{\bibfnamefont{J.~L.} \bibnamefont{Bohn}}, \bibnamefont{and}
  \bibinfo{author}{\bibfnamefont{J.}~\bibnamefont{Ye}},
  \bibinfo{journal}{Nature} \textbf{\bibinfo{volume}{492}},
  \bibinfo{pages}{396} (\bibinfo{year}{2012}).

\bibitem[{\citenamefont{Hummon et~al.}(2013)\citenamefont{Hummon, Yeo, Stuhl,
  Collopy, Xia, and Ye}}]{Hummon2013}
\bibinfo{author}{\bibfnamefont{M.~T.} \bibnamefont{Hummon}},
  \bibinfo{author}{\bibfnamefont{M.}~\bibnamefont{Yeo}},
  \bibinfo{author}{\bibfnamefont{B.~K.} \bibnamefont{Stuhl}},
  \bibinfo{author}{\bibfnamefont{A.~L.} \bibnamefont{Collopy}},
  \bibinfo{author}{\bibfnamefont{Y.}~\bibnamefont{Xia}}, \bibnamefont{and}
  \bibinfo{author}{\bibfnamefont{J.}~\bibnamefont{Ye}}, \bibinfo{journal}{Phys.
  Rev. Lett.} \textbf{\bibinfo{volume}{110}}, \bibinfo{pages}{143001}
  (\bibinfo{year}{2013}).

\bibitem[{\citenamefont{Staanum et~al.}(2010)\citenamefont{Staanum, Højbjerre,
  Skyt, Hansen, and Drewsen}}]{Staanum2010}
\bibinfo{author}{\bibfnamefont{P.~F.} \bibnamefont{Staanum}},
  \bibinfo{author}{\bibfnamefont{K.}~\bibnamefont{Højbjerre}},
  \bibinfo{author}{\bibfnamefont{P.~S.} \bibnamefont{Skyt}},
  \bibinfo{author}{\bibfnamefont{A.~K.} \bibnamefont{Hansen}},
  \bibnamefont{and} \bibinfo{author}{\bibfnamefont{M.}~\bibnamefont{Drewsen}},
  \bibinfo{journal}{Nat. Phys.} \textbf{\bibinfo{volume}{6}},
  \bibinfo{pages}{271} (\bibinfo{year}{2010}).

\bibitem[{\citenamefont{Nguyen et~al.}(2011)\citenamefont{Nguyen, Viteri,
  Hohenstein, Sherrill, Brown, and Odom}}]{Nguyen2011}
\bibinfo{author}{\bibfnamefont{J.~H.~V.} \bibnamefont{Nguyen}},
  \bibinfo{author}{\bibfnamefont{C.~R.} \bibnamefont{Viteri}},
  \bibinfo{author}{\bibfnamefont{E.~G.} \bibnamefont{Hohenstein}},
  \bibinfo{author}{\bibfnamefont{C.~D.} \bibnamefont{Sherrill}},
  \bibinfo{author}{\bibfnamefont{K.~R.} \bibnamefont{Brown}}, \bibnamefont{and}
  \bibinfo{author}{\bibfnamefont{B.}~\bibnamefont{Odom}}, \bibinfo{journal}{New
  J. Phys.} \textbf{\bibinfo{volume}{13}}, \bibinfo{pages}{063023}
  (\bibinfo{year}{2011}).

\bibitem[{\citenamefont{Bressel et~al.}(2012)\citenamefont{Bressel, Borodin,
  Shen, Hansen, Ernsting, and Schiller}}]{Bressel2012}
\bibinfo{author}{\bibfnamefont{U.}~\bibnamefont{Bressel}},
  \bibinfo{author}{\bibfnamefont{A.}~\bibnamefont{Borodin}},
  \bibinfo{author}{\bibfnamefont{J.}~\bibnamefont{Shen}},
  \bibinfo{author}{\bibfnamefont{M.}~\bibnamefont{Hansen}},
  \bibinfo{author}{\bibfnamefont{I.}~\bibnamefont{Ernsting}}, \bibnamefont{and}
  \bibinfo{author}{\bibfnamefont{S.}~\bibnamefont{Schiller}},
  \bibinfo{journal}{Phys. Rev. Lett.} \textbf{\bibinfo{volume}{108}},
  \bibinfo{pages}{183003} (\bibinfo{year}{2012}).

\bibitem[{\citenamefont{Willitsch
  et~al.}(2008{\natexlab{a}})\citenamefont{Willitsch, Bell, Gingell, and
  Softley}}]{Willitsch2008a}
\bibinfo{author}{\bibfnamefont{S.}~\bibnamefont{Willitsch}},
  \bibinfo{author}{\bibfnamefont{M.~T.} \bibnamefont{Bell}},
  \bibinfo{author}{\bibfnamefont{A.~D.} \bibnamefont{Gingell}},
  \bibnamefont{and} \bibinfo{author}{\bibfnamefont{T.~P.}
  \bibnamefont{Softley}}, \bibinfo{journal}{Phys. Chem. Chem. Phys.}
  \textbf{\bibinfo{volume}{10}}, \bibinfo{pages}{7200}
  (\bibinfo{year}{2008}{\natexlab{a}}).

\bibitem[{\citenamefont{Meyer and Bohn}(2008)}]{Meyer2008}
\bibinfo{author}{\bibfnamefont{E.~R.} \bibnamefont{Meyer}} \bibnamefont{and}
  \bibinfo{author}{\bibfnamefont{J.~L.} \bibnamefont{Bohn}},
  \bibinfo{journal}{Phys. Rev. A} \textbf{\bibinfo{volume}{78}},
  \bibinfo{pages}{010502} (\bibinfo{year}{2008}).

\bibitem[{\citenamefont{Cossel et~al.}(2012)\citenamefont{Cossel, Gresh,
  Sinclair, Coffey, Skripnikov, Petrov, Mosyagin, Titov, Field, Meyer
  et~al.}}]{Cossel2012}
\bibinfo{author}{\bibfnamefont{K.~C.} \bibnamefont{Cossel}},
  \bibinfo{author}{\bibfnamefont{D.~N.} \bibnamefont{Gresh}},
  \bibinfo{author}{\bibfnamefont{L.~C.} \bibnamefont{Sinclair}},
  \bibinfo{author}{\bibfnamefont{T.}~\bibnamefont{Coffey}},
  \bibinfo{author}{\bibfnamefont{L.~V.} \bibnamefont{Skripnikov}},
  \bibinfo{author}{\bibfnamefont{A.~N.} \bibnamefont{Petrov}},
  \bibinfo{author}{\bibfnamefont{N.~S.} \bibnamefont{Mosyagin}},
  \bibinfo{author}{\bibfnamefont{A.~V.} \bibnamefont{Titov}},
  \bibinfo{author}{\bibfnamefont{R.~W.} \bibnamefont{Field}},
  \bibinfo{author}{\bibfnamefont{E.~R.} \bibnamefont{Meyer}},
  \bibnamefont{et~al.}, \bibinfo{journal}{Chem. Phys. Lett.}
  \textbf{\bibinfo{volume}{546}}, \bibinfo{pages}{1} (\bibinfo{year}{2012}).

\bibitem[{\citenamefont{Flambaum and Kozlov}(2007)}]{Flambaum2007}
\bibinfo{author}{\bibfnamefont{V.~V.} \bibnamefont{Flambaum}} \bibnamefont{and}
  \bibinfo{author}{\bibfnamefont{M.~G.} \bibnamefont{Kozlov}},
  \bibinfo{journal}{Phys. Rev. Lett.} \textbf{\bibinfo{volume}{99}},
  \bibinfo{pages}{150801} (\bibinfo{year}{2007}).

\bibitem[{\citenamefont{Schuster et~al.}(2011)\citenamefont{Schuster, Bishop,
  Chuang, {DeMille}, and Schoelkopf}}]{Schuster2011}
\bibinfo{author}{\bibfnamefont{D.~I.} \bibnamefont{Schuster}},
  \bibinfo{author}{\bibfnamefont{L.~S.} \bibnamefont{Bishop}},
  \bibinfo{author}{\bibfnamefont{I.~L.} \bibnamefont{Chuang}},
  \bibinfo{author}{\bibfnamefont{D.}~\bibnamefont{{DeMille}}},
  \bibnamefont{and} \bibinfo{author}{\bibfnamefont{R.~J.}
  \bibnamefont{Schoelkopf}}, \bibinfo{journal}{Phys. Rev. A}
  \textbf{\bibinfo{volume}{83}}, \bibinfo{pages}{012311}
  (\bibinfo{year}{2011}).

\bibitem[{\citenamefont{Hudson}(2009)}]{Hudson2009}
\bibinfo{author}{\bibfnamefont{E.~R.} \bibnamefont{Hudson}},
  \bibinfo{journal}{Phys. Rev. A} \textbf{\bibinfo{volume}{79}},
  \bibinfo{pages}{032716} (\bibinfo{year}{2009}).

\bibitem[{\citenamefont{Rellergert et~al.}(2013)\citenamefont{Rellergert,
  Sullivan, Schowalter, Kotochigova, Chen, and Hudson}}]{Rellergert2013}
\bibinfo{author}{\bibfnamefont{W.~G.} \bibnamefont{Rellergert}},
  \bibinfo{author}{\bibfnamefont{S.~T.} \bibnamefont{Sullivan}},
  \bibinfo{author}{\bibfnamefont{S.~J.} \bibnamefont{Schowalter}},
  \bibinfo{author}{\bibfnamefont{S.}~\bibnamefont{Kotochigova}},
  \bibinfo{author}{\bibfnamefont{K.}~\bibnamefont{Chen}}, \bibnamefont{and}
  \bibinfo{author}{\bibfnamefont{E.~R.} \bibnamefont{Hudson}},
  \bibinfo{journal}{Nature} \textbf{\bibinfo{volume}{495}},
  \bibinfo{pages}{490} (\bibinfo{year}{2013}).

\bibitem[{\citenamefont{Chen et~al.}(2011)\citenamefont{Chen, Schowalter,
  Kotochigova, Petrov, Rellergert, Sullivan, and Hudson}}]{Chen2011}
\bibinfo{author}{\bibfnamefont{K.}~\bibnamefont{Chen}},
  \bibinfo{author}{\bibfnamefont{S.~J.} \bibnamefont{Schowalter}},
  \bibinfo{author}{\bibfnamefont{S.}~\bibnamefont{Kotochigova}},
  \bibinfo{author}{\bibfnamefont{A.}~\bibnamefont{Petrov}},
  \bibinfo{author}{\bibfnamefont{W.~G.} \bibnamefont{Rellergert}},
  \bibinfo{author}{\bibfnamefont{S.~T.} \bibnamefont{Sullivan}},
  \bibnamefont{and} \bibinfo{author}{\bibfnamefont{E.~R.}
  \bibnamefont{Hudson}}, \bibinfo{journal}{Phys. Rev. A}
  \textbf{\bibinfo{volume}{83}}, \bibinfo{pages}{030501}
  (\bibinfo{year}{2011}).

\bibitem[{\citenamefont{Schowalter et~al.}(2012)\citenamefont{Schowalter, Chen,
  Rellergert, Sullivan, and Hudson}}]{Schowalter2012}
\bibinfo{author}{\bibfnamefont{S.~J.} \bibnamefont{Schowalter}},
  \bibinfo{author}{\bibfnamefont{K.}~\bibnamefont{Chen}},
  \bibinfo{author}{\bibfnamefont{W.~G.} \bibnamefont{Rellergert}},
  \bibinfo{author}{\bibfnamefont{S.~T.} \bibnamefont{Sullivan}},
  \bibnamefont{and} \bibinfo{author}{\bibfnamefont{E.~R.}
  \bibnamefont{Hudson}}, \bibinfo{journal}{Rev. Sci. Instrum.}
  \textbf{\bibinfo{volume}{83}}, \bibinfo{pages}{043103}
  (\bibinfo{year}{2012}).

\bibitem[{\citenamefont{Harvey et~al.}(1997)\citenamefont{Harvey, Schröder,
  Koch, Danovich, Shaik, and Schwarz}}]{Harvey1997}
\bibinfo{author}{\bibfnamefont{J.~N.} \bibnamefont{Harvey}},
  \bibinfo{author}{\bibfnamefont{D.}~\bibnamefont{Schröder}},
  \bibinfo{author}{\bibfnamefont{W.}~\bibnamefont{Koch}},
  \bibinfo{author}{\bibfnamefont{D.}~\bibnamefont{Danovich}},
  \bibinfo{author}{\bibfnamefont{S.}~\bibnamefont{Shaik}}, \bibnamefont{and}
  \bibinfo{author}{\bibfnamefont{H.}~\bibnamefont{Schwarz}},
  \bibinfo{journal}{Chem. Phys. Lett.} \textbf{\bibinfo{volume}{273}},
  \bibinfo{pages}{164} (\bibinfo{year}{1997}).

\bibitem[{\citenamefont{Willitsch
  et~al.}(2008{\natexlab{b}})\citenamefont{Willitsch, Bell, Gingell, Procter,
  and Softley}}]{Willitsch2008}
\bibinfo{author}{\bibfnamefont{S.}~\bibnamefont{Willitsch}},
  \bibinfo{author}{\bibfnamefont{M.~T.} \bibnamefont{Bell}},
  \bibinfo{author}{\bibfnamefont{A.~D.} \bibnamefont{Gingell}},
  \bibinfo{author}{\bibfnamefont{S.~R.} \bibnamefont{Procter}},
  \bibnamefont{and} \bibinfo{author}{\bibfnamefont{T.~P.}
  \bibnamefont{Softley}}, \bibinfo{journal}{Phys. Rev. Lett.}
  \textbf{\bibinfo{volume}{100}}, \bibinfo{pages}{043203}
  (\bibinfo{year}{2008}{\natexlab{b}}).

\bibitem[{\citenamefont{Gingell et~al.}(2010)\citenamefont{Gingell, Bell,
  Oldham, Softley, and Harvey}}]{Gingell2010}
\bibinfo{author}{\bibfnamefont{A.~D.} \bibnamefont{Gingell}},
  \bibinfo{author}{\bibfnamefont{M.~T.} \bibnamefont{Bell}},
  \bibinfo{author}{\bibfnamefont{J.~M.} \bibnamefont{Oldham}},
  \bibinfo{author}{\bibfnamefont{T.~P.} \bibnamefont{Softley}},
  \bibnamefont{and} \bibinfo{author}{\bibfnamefont{J.~N.}
  \bibnamefont{Harvey}}, \bibinfo{journal}{J. Chem. Phys.}
  \textbf{\bibinfo{volume}{133}}, \bibinfo{pages}{194302}
  (\bibinfo{year}{2010}).

\bibitem[{\citenamefont{Churchill et~al.}(2011)\citenamefont{Churchill,
  {DePalatis}, and Chapman}}]{Churchill2011}
\bibinfo{author}{\bibfnamefont{L.~R.} \bibnamefont{Churchill}},
  \bibinfo{author}{\bibfnamefont{M.~V.} \bibnamefont{{DePalatis}}},
  \bibnamefont{and} \bibinfo{author}{\bibfnamefont{M.~S.}
  \bibnamefont{Chapman}}, \bibinfo{journal}{Phys. Rev. A}
  \textbf{\bibinfo{volume}{83}}, \bibinfo{pages}{012710}
  (\bibinfo{year}{2011}).

\bibitem[{\citenamefont{Rohde et~al.}(2010)\citenamefont{Rohde, Almendros,
  Schuck, Huwer, Hennrich, and Eschner}}]{Rohde2010}
\bibinfo{author}{\bibfnamefont{F.}~\bibnamefont{Rohde}},
  \bibinfo{author}{\bibfnamefont{M.}~\bibnamefont{Almendros}},
  \bibinfo{author}{\bibfnamefont{C.}~\bibnamefont{Schuck}},
  \bibinfo{author}{\bibfnamefont{J.}~\bibnamefont{Huwer}},
  \bibinfo{author}{\bibfnamefont{M.}~\bibnamefont{Hennrich}}, \bibnamefont{and}
  \bibinfo{author}{\bibfnamefont{J.}~\bibnamefont{Eschner}},
  \bibinfo{journal}{J. Phys. B: At., Mol. Opt. Phys.}
  \textbf{\bibinfo{volume}{43}}, \bibinfo{pages}{115401}
  (\bibinfo{year}{2010}).

\bibitem[{\citenamefont{Lide}(2007)}]{Lide2007}
\bibinfo{author}{\bibfnamefont{D.~R.} \bibnamefont{Lide}},
  \emph{\bibinfo{title}{{CRC} Handbook of Chemistry and Physics}}
  (\bibinfo{publisher}{{CRC} Press}, \bibinfo{address}{Boca Raton, {FL}},
  \bibinfo{year}{2007}), \bibinfo{edition}{88th} ed.

\bibitem[{\citenamefont{Gioumousis and Stevenson}(1958)}]{Gioumousis1958}
\bibinfo{author}{\bibfnamefont{G.}~\bibnamefont{Gioumousis}} \bibnamefont{and}
  \bibinfo{author}{\bibfnamefont{D.~P.} \bibnamefont{Stevenson}},
  \bibinfo{journal}{J. Chem. Phys.} \textbf{\bibinfo{volume}{29}},
  \bibinfo{pages}{294} (\bibinfo{year}{1958}).

\bibitem[{\citenamefont{Troe}(1985)}]{Troe1985}
\bibinfo{author}{\bibfnamefont{J.}~\bibnamefont{Troe}}, \bibinfo{journal}{Chem.
  Phys. Lett.} \textbf{\bibinfo{volume}{122}}, \bibinfo{pages}{425}
  (\bibinfo{year}{1985}).

\bibitem[{\citenamefont{Su and Bowers}(1973)}]{Su1973}
\bibinfo{author}{\bibfnamefont{T.}~\bibnamefont{Su}} \bibnamefont{and}
  \bibinfo{author}{\bibfnamefont{M.~T.} \bibnamefont{Bowers}},
  \bibinfo{journal}{J. Chem. Phys} \textbf{\bibinfo{volume}{58}},
  \bibinfo{pages}{3027} (\bibinfo{year}{1973}).

\bibitem[{Var(2004)}]{Varian2004}
\emph{\bibinfo{title}{UHV-24/UHV-24p Ionization Gauge Instruction Manual}},
  \bibinfo{organization}{Agilent Technologies}, \bibinfo{edition}{{Rev. E}} ed.
  (\bibinfo{year}{2004}).

\bibitem[{\citenamefont{Karpas and Berant}(1989)}]{Karpas1989}
\bibinfo{author}{\bibfnamefont{Z.}~\bibnamefont{Karpas}} \bibnamefont{and}
  \bibinfo{author}{\bibfnamefont{Z.}~\bibnamefont{Berant}},
  \bibinfo{journal}{J. Phys. Chem.} \textbf{\bibinfo{volume}{93}},
  \bibinfo{pages}{3021} (\bibinfo{year}{1989}).

\bibitem[{\citenamefont{Olney et~al.}(1997)\citenamefont{Olney, Cann, Cooper,
  and Brion}}]{Olney1997}
\bibinfo{author}{\bibfnamefont{T.~N.} \bibnamefont{Olney}},
  \bibinfo{author}{\bibfnamefont{N.}~\bibnamefont{Cann}},
  \bibinfo{author}{\bibfnamefont{G.}~\bibnamefont{Cooper}}, \bibnamefont{and}
  \bibinfo{author}{\bibfnamefont{C.}~\bibnamefont{Brion}},
  \bibinfo{journal}{Chem. Phys.} \textbf{\bibinfo{volume}{223}},
  \bibinfo{pages}{59} (\bibinfo{year}{1997}).

\bibitem[{\citenamefont{Maroulis et~al.}(1997)\citenamefont{Maroulis, Makris,
  Hohm, and Goebel}}]{Maroulis1997}
\bibinfo{author}{\bibfnamefont{G.}~\bibnamefont{Maroulis}},
  \bibinfo{author}{\bibfnamefont{C.}~\bibnamefont{Makris}},
  \bibinfo{author}{\bibfnamefont{U.}~\bibnamefont{Hohm}}, \bibnamefont{and}
  \bibinfo{author}{\bibfnamefont{D.}~\bibnamefont{Goebel}},
  \bibinfo{journal}{J. Phys. Chem. A} \textbf{\bibinfo{volume}{101}},
  \bibinfo{pages}{953} (\bibinfo{year}{1997}).

\bibitem[{\citenamefont{Maroulis and Makris}(1997)}]{Maroulis1997a}
\bibinfo{author}{\bibfnamefont{G.}~\bibnamefont{Maroulis}} \bibnamefont{and}
  \bibinfo{author}{\bibfnamefont{C.}~\bibnamefont{Makris}},
  \bibinfo{journal}{Mol. Phys.} \textbf{\bibinfo{volume}{91}},
  \bibinfo{pages}{333} (\bibinfo{year}{1997}).

\bibitem[{\citenamefont{{R. D. Johnson, ed.}}(2011)}]{Johnson2011}
\bibinfo{author}{\bibnamefont{{R. D. Johnson, ed.}}},
  \emph{\bibinfo{title}{{NIST} computational chemistry comparison and benchmark
  database}} (\bibinfo{year}{2011}), \bibinfo{note}{release 15b},
  \urlprefix\url{http://cccbdb.nist.gov/}.

\bibitem[{\citenamefont{Roth et~al.}(2005)\citenamefont{Roth, Ostendorf, Wenz,
  and Schiller}}]{Roth2005}
\bibinfo{author}{\bibfnamefont{B.}~\bibnamefont{Roth}},
  \bibinfo{author}{\bibfnamefont{A.}~\bibnamefont{Ostendorf}},
  \bibinfo{author}{\bibfnamefont{H.}~\bibnamefont{Wenz}}, \bibnamefont{and}
  \bibinfo{author}{\bibfnamefont{S.}~\bibnamefont{Schiller}},
  \bibinfo{journal}{J. Phys. B: At., Mol. Opt. Phys.}
  \textbf{\bibinfo{volume}{38}}, \bibinfo{pages}{3673} (\bibinfo{year}{2005}).

\bibitem[{\citenamefont{Roth et~al.}(2007)\citenamefont{Roth, Blythe, and
  Schiller}}]{Roth2007}
\bibinfo{author}{\bibfnamefont{B.}~\bibnamefont{Roth}},
  \bibinfo{author}{\bibfnamefont{P.}~\bibnamefont{Blythe}}, \bibnamefont{and}
  \bibinfo{author}{\bibfnamefont{S.}~\bibnamefont{Schiller}},
  \bibinfo{journal}{Phys. Rev. A} \textbf{\bibinfo{volume}{75}},
  \bibinfo{pages}{023402} (\bibinfo{year}{2007}).

\bibitem[{\citenamefont{Landa et~al.}(2012)\citenamefont{Landa, Drewsen,
  Reznik, and Retzker}}]{Landa2012}
\bibinfo{author}{\bibfnamefont{H.}~\bibnamefont{Landa}},
  \bibinfo{author}{\bibfnamefont{M.}~\bibnamefont{Drewsen}},
  \bibinfo{author}{\bibfnamefont{B.}~\bibnamefont{Reznik}}, \bibnamefont{and}
  \bibinfo{author}{\bibfnamefont{A.}~\bibnamefont{Retzker}},
  \bibinfo{journal}{New J. Phys.} \textbf{\bibinfo{volume}{14}},
  \bibinfo{pages}{093023} (\bibinfo{year}{2012}).

\bibitem[{\citenamefont{Baba and Waki}(2002)}]{Baba2002}
\bibinfo{author}{\bibfnamefont{T.}~\bibnamefont{Baba}} \bibnamefont{and}
  \bibinfo{author}{\bibfnamefont{I.}~\bibnamefont{Waki}}, \bibinfo{journal}{J.
  Appl. Phys.} \textbf{\bibinfo{volume}{92}}, \bibinfo{pages}{4109}
  (\bibinfo{year}{2002}).

\bibitem[{\citenamefont{Lias}(2011)}]{Lias2011}
\bibinfo{author}{\bibfnamefont{S.~G.} \bibnamefont{Lias}}, in
  \emph{\bibinfo{booktitle}{NIST Chemistry WebBook}}, edited by
  \bibinfo{editor}{\bibfnamefont{P.~J.} \bibnamefont{Linstrom}}
  \bibnamefont{and} \bibinfo{editor}{\bibfnamefont{W.~G.}
  \bibnamefont{Mallard}} (\bibinfo{publisher}{National Institute of Standards
  and Technology}, \bibinfo{address}{Gaithersburg MD, 20899},
  \bibinfo{year}{2011}), \urlprefix\url{http://webbook.nist.gov}.

\bibitem[{\citenamefont{Karlsson and Litzén}(1999)}]{Karlsson1999}
\bibinfo{author}{\bibfnamefont{H.}~\bibnamefont{Karlsson}} \bibnamefont{and}
  \bibinfo{author}{\bibfnamefont{U.}~\bibnamefont{Litzén}},
  \bibinfo{journal}{Phys. Scr.} \textbf{\bibinfo{volume}{60}},
  \bibinfo{pages}{321} (\bibinfo{year}{1999}).

\bibitem[{\citenamefont{Tachikawa}(2000)}]{Tachikawa2000}
\bibinfo{author}{\bibfnamefont{H.}~\bibnamefont{Tachikawa}},
  \bibinfo{journal}{J. Phys. B: At., Mol. Opt. Phys.}
  \textbf{\bibinfo{volume}{33}}, \bibinfo{pages}{1725} (\bibinfo{year}{2000}).

\bibitem[{\citenamefont{Murad}(1982)}]{Murad1982}
\bibinfo{author}{\bibfnamefont{E.}~\bibnamefont{Murad}}, \bibinfo{journal}{J.
  Chem. Phys.} \textbf{\bibinfo{volume}{77}}, \bibinfo{pages}{2057}
  (\bibinfo{year}{1982}).

\bibitem[{\citenamefont{Roth et~al.}(2008)\citenamefont{Roth, Offenberg, Zhang,
  and Schiller}}]{Roth2008}
\bibinfo{author}{\bibfnamefont{B.}~\bibnamefont{Roth}},
  \bibinfo{author}{\bibfnamefont{D.}~\bibnamefont{Offenberg}},
  \bibinfo{author}{\bibfnamefont{C.~B.} \bibnamefont{Zhang}}, \bibnamefont{and}
  \bibinfo{author}{\bibfnamefont{S.}~\bibnamefont{Schiller}},
  \bibinfo{journal}{Phys. Rev. A} \textbf{\bibinfo{volume}{78}},
  \bibinfo{pages}{042709} (\bibinfo{year}{2008}).

\bibitem[{\citenamefont{Drewsen et~al.}(2004)\citenamefont{Drewsen, Mortensen,
  Martinussen, Staanum, and Sørensen}}]{Drewsen2004}
\bibinfo{author}{\bibfnamefont{M.}~\bibnamefont{Drewsen}},
  \bibinfo{author}{\bibfnamefont{A.}~\bibnamefont{Mortensen}},
  \bibinfo{author}{\bibfnamefont{R.}~\bibnamefont{Martinussen}},
  \bibinfo{author}{\bibfnamefont{P.}~\bibnamefont{Staanum}}, \bibnamefont{and}
  \bibinfo{author}{\bibfnamefont{J.~L.} \bibnamefont{Sørensen}},
  \bibinfo{journal}{Phys. Rev. Lett.} \textbf{\bibinfo{volume}{93}},
  \bibinfo{pages}{243201} (\bibinfo{year}{2004}).

\end{thebibliography}

\end{document}